\documentstyle[prl,aps,epsf,multicol]{revtex}

\begin{document}
\title{Nature of mechanical instabilities and their effect on kinetic friction}

\author{Martin H. M\"user}

\address{Institut f\"ur Physik, WA 331; Johannes Gutenberg Universit\"at;
         55099 Mainz; Germany}

\date{\today}
\maketitle

\begin{abstract}
It has long been recognized that the key to understand kinetic friction 
force $F_k$ 
is the analysis of microscopic instabilities that lead to sudden irreversible 
"pops" of certain degrees of freedom. In this Letter, the nature of such 
instabilities is characterized with an emphasis on boundary lubricants. It is 
shown that there are certain critical values of the parameters defining our 
model Hamiltonian, where the behavior of the instabilities changes 
qualitatively. Simultaneously, the  functional dependence of $F_k$ on the 
sliding velocity $v_0$ changes.  The relevant parameters studied here are 
dimensionality of the interface, degree of commensurability, first higher 
harmonic in the lubricant wall potential, and temperature. Molecular dynamics 
simulations are carried out to test whether the predictions made on the basis 
of the simple model also hold in less  idealized circumstances.
\end{abstract}

\begin{multicols}{2}

When a solid slider is moved laterally with respect to a substrate, the kinetic 
friction $F_k$ is usually almost independent of the sliding velocity 
$v_0$~\cite{dowson79} with leading corrections in the order of 
$\ln v_0$~\cite{dieterich79}. This so-called Coulomb friction 
differs from drag or Stokes friction that states a linear relation 
between  $F_k$ and $v_0$ and that can be understood from 
equilibrium  statistical mechanics as a  special  case of the fluctuation 
dissipation theorem~\cite{hansen86}:  
For instance, the drag force  experienced  by a Brownian 
particle in solution  arises from the many collisions between  the Brownian 
particle and 
the solvent molecules. In general, the condition for a linear friction  force 
to occur is the  (anharmonic) coupling of a central degree of freedom like the  
Brownian particle  or a phonon to (infinitely) many other discrete degrees of 
freedom.  In this  Letter, the question will be addressed what the condition is 
for Coulomb  friction to occur.

It has long been recognized that Coulomb friction must be related to 
instabilities  that occur on a microscopic 
scale~\cite{brillouin09}. When a slider is moved 
laterally with respect to the substrate, fast motion (pops) of certain degrees 
of freedom become  unavoidable even if the slider´s center of mass velocity is 
extremely small. The microscopic peak velocities in such pops are rather 
independent of $v_0$ and  consequently the energy dissipated via a Stokes-type 
mechanism also becomes almost  independent of $v_0$. The prototypical 
instability  leading to Coulomb friction  was suggested by Prandtl and 
Tomlinson~\cite{prandtl28}. 
In their one-dimensional model, a surface atom in the slider is  
coupled with a spring of stiffness $k$ to its ideal  lattice site  
which moves at constant  velocity $v_0$. Interactions with the rigid
substrate are modeled with a  potential energy surface $V$ 
that is periodic in the  substrate's lattice constant plus some drag force 
linear in  the atom's veloctity $\dot{x}$. If $k$ is sufficiently small, pops 
become unavoidable and  neglecting thermal fluctuations, $F_k$ remains finite 
in the limit of zero $v_0$.  Instabilities and small-velocity kinetic friction 
in more complex model systems  have been investigated since the Prandtl 
Tomlinson model was introduced. In particular,  the $F_k(v_0)$ 
relationship of elastic manifolds sliding in ordered and disordered media has 
been studied extensively~\cite{fisher85,chauve01}.
The functional dependence is commonly found to be

\begin{equation}
F_k(v) = F_k(0) + c v^\beta.
\label{eq:fk_athermal}
\end{equation}

While the elastic instabilities considered in those models are important in 
various contexts, this does not seem to be the case for the atomistic 
explanation of solid friction: Many detailed calculations and atomistic 
computer simulations reveal that in almost all cases, interbulk interactions 
are too weak to lead to instabilities at the atomic scale and as a 
consequence $F_k(0) = 0$~\cite{robbins01}. 
If interbulk interactions are very strong, irreversible processes 
like plastic deformation, material mixing, cold-welding, etc. usually occur and 
prevent the instabilities from being elastic.

It has been suggested that the presence of adsorbed particles, i.e. a boundary
lubricant, confined between two surfaces is a more likely explanation for the 
commonly observed presence of solid friction~\cite{he99}. 
The main argument is that 
molecules that are only weakly bound to either surface can accommodate the 
surface corrugation of both walls simultaneously, which locks the walls 
together. This argument leads to static friction, which is the minimum force to 
initiate sliding between two  solids. However, in order to explain kinetic 
friction, it is necessary to analyze  the adiabatic solution 
$x_{\rm ad}(t)$ of the lubricant atoms. 
$x_{\rm ad}(t)$ is the athermal trajectory of an atom that always
relaxes to the closest mechanical equilibrium position at every
instant of time.
The analysis of $x_{\rm ad}(t)$ has  become common 
practice in the context of the motion of elastic manifolds in  
disordered media~\cite{fisher85}. 
In the impurity limit, interactions between lubricant  atoms can be 
neglected and consequently $x_{\rm ad}(t)$ merely depends on the
initial condition and the relative motion of top and bottom wall.
In the following, $x_{\rm ad}(t)$ and its  connection to 
kinetic friction will be examined. 
To the best of the author's knowledge such an analysis 
has not yet been done, although the present model has already been used 
extensively to predict successfully various tribological 
phenomena~\cite{rozman96,rozman98}.

The equation of motion for a lubricant atom in the boundary regime reads:
\begin{eqnarray}
\ddot{x} & = & -\gamma_b \dot{x}   - \gamma_t (\dot{x} - v_0)
                + {1\over m} \Gamma(t) \nonumber\\
          &   & - {1\over m} {\partial \over \partial x } 
                \left\{ V_b(x) + V_t(x-v_0t) \right \},
\label{eq:eq_of_motion}
\end{eqnarray}
where $x$ denotes the atom's position, $m$ is the atom's mass, $\gamma_t$ and 
$\gamma_b$ parametrize the damping forces from the top and the bottom wall, and 
$V_t$ and $V_b$ denote the interaction of  the confined atom with slider and 
the substrate. $\Gamma(t)$ is a Langevin type stochastic random force defining
temperature. For the lubricant wall interactions, various choices will be 
considered. Centrosymmetric choices can all be written in the form:
\begin{equation}
V_{t,b} = V_{t,b}^{(0)} \cos( x/b_{t,b}) 
        + V_{t,b}^{(1)} \cos(2  x/b_{t,b}) + ...
\label{eq:potential}
\end{equation}
where $2\pi b_t$ and $2\pi b_b$ are the period of the top and the bottom wall,
respectively. The relevant physical units are defined through the choice
$V_t^{(0)} = 1$, $m = 1$, $b_t =1$, and Boltzmann's constant $k_B = 1$. 
Furthermore, we will only consider slightly underdamped dynamics 
($\gamma_b = 1 \Rightarrow x(t) \approx x_{\rm ad}(t)$) and 
restrict ourselves to the symmetric choice of 
$V_0 := V_{t,b}^{(0)}$. The  free parameters are thus the degree of 
lattice mismatch ($b_t-b_b$), the value of the first higher 
harmonic ($V_1 = V_{t,b}^{(1)}$), and temperature $T$.

\begin{figure}[hbtp]
\begin{center} \leavevmode \vspace*{+2mm}
 \hbox{ \epsfxsize=75mm
 \epsfbox{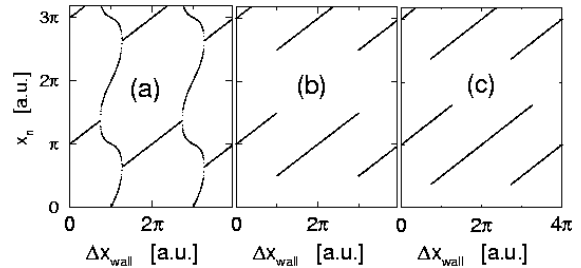}}
 \begin{minipage}{8.5cm}
 \caption{Mechanical equilibrium positions for adsorbed atoms between
 two commensurate solid surfaces as a function of the relative displacement
 $\Delta x_{\rm wall}$ between the walls. The grey lines indicates
 the solutions of 
 Eqs.~(\protect\ref{eq:eq_of_motion}),(\protect\ref{eq:potential}) 
 if the walls are in slow relative sliding motion. 
 (a) First higher harmonic $V_1 < 0$. (b) $V_1 = 0$. (c) $V_1 > 0$.}
 \label{fig:triptich}
 \end{minipage}
\end{center}
\end{figure}

We start the discussion of a commensurate (com.) system ($b_t = b_b$)
without higher harmonics in the absence of thermal fluctuations. In that case,
the net time-dependent potential is simply given by 
$V(t) = 2 V_0 \cos\left({1\over 2b} vt\right)
               \cos\left[{1\over b} (x-vt/2)\right]$.
Thus, for times $\cos\left({1\over 2b} vt\right) \ne 0$, 
the atoms move at velocity 
$v_0/2$ as shown in Fig.~\ref{fig:triptich}. 
An infinitely small moment after a situation where 
this inequality was not satisfied, an atom will not be able to find a stable 
position in the immediate vicinity of the previous stable position. As the 
slider moves on with respect to the substrate, the atoms should therefore slide 
rapidly towards a new mechanical equilibrium. However, in order for this to 
happen, one needs a symmetry breaking element such as thermal fluctuations, 
$\gamma_b \ne \gamma_t$, $b_t \ne b_b$, or round-off errors in numerical
calculations.
Otherwise all stable trajectories simply are $x = n \pi + v_0 t/2$ where $n$ 
is an integer number and the average (conservative) force of the lubricant atom 
acting on either wall will average to zero. 
For the athermal com. system, we chose to break symmetry by violating
Galilei invariance and set $\gamma_t = 0$. 
This choice is arbitrary but convenient, and is a better-controlled
procedure than relying on round-off erros. We note also that for inc. walls,
there is no physical reason for assuming the equality $\gamma_t = \gamma_b$.

Owing to broken symmetry ($\gamma_t =0$), the atom can now slide to the next 
minimum indicated by the gray lines in Fig.~\ref{fig:triptich}b. 
However, the peak 
velocities $\dot{x}_p$ in this process remain much smaller than 
during a pop within the Prandtl Tomlinson model. The reason is that due to the 
symmetry of $V_t$ and  $V_b$, the pops occur between equivalent positions in 
the limit of arbitrarily small $v_0$. 
Since no lower bound for the dissipated energy can be 
given, the zero-velocity $F_k$  will be zero. At the same time, there
is no upper bound for the ratio $\dot{x}_p/v_0$, so that 
$F_k$ cannot 
simply vanish linearly with $v_0$ but only with some power $\beta$ 
smaller than unity.

\begin{figure}[hbtp]
\begin{center} \leavevmode
 \hbox{ \epsfxsize=70mm
 \epsfbox{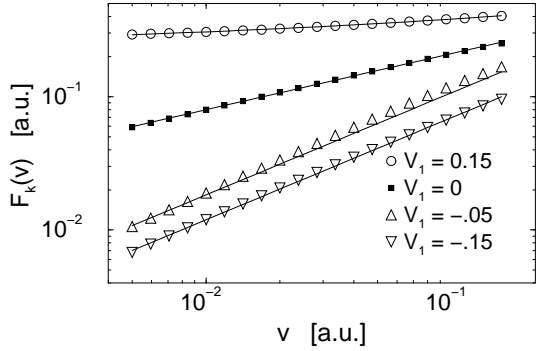} }
 \begin{minipage}{8.5cm}
 \caption{Kinetic friction force $F_k$ as a function  of sliding
 velocity $v_0$ for commensurate walls with different first higher
 harmonics. Straight lines are fits to low $v_0$ data according to
 Eq.~(\protect\ref{eq:fk_athermal}). The two data sets with $V_1 < 0$
 are fitted with the same exponent $\beta = 0.74(0)$.
 }
 \label{fig:force_comm}
 \end{minipage}
\end{center}
\end{figure}

The nature of the instabilities 
changes qualitatively when the first higher 
harmonic is different from zero. Hence, in the sense of 
Morse theory~\cite{morse63}, which 
contains Landau's theory of phase transition as a special case, 
the com. system without higher harmonics can be considered a 
(multi-/tri-) critical point. If $V_1 < 0$, the adiabatic solution $x(t)$ 
becomes continuous as shown in Fig.~\ref{fig:triptich}, 
however, the time derivative $\dot{x}(t)$ diverges. If $V_1 > 0$, the adiabatic 
solution is discontinuous and the pops occur between inequivalent positions 
as shown in Fig.~\ref{fig:triptich}c. 
In analogy to phase transitions,  the pops occurring 
for $V_1 > 0$ shall be called first-order instabilities, those for 
$V_1 \le 0$ second-order 
instabilities. Only for first-order instabilities can one expect finite energy 
dissipation and thus finite $F_k$ as $v_0$ approaches zero. 
A numerical analysis shows that all three 
cases can be described with Eq.~(\ref{eq:fk_athermal}). 
The results are shown in Fig.~\ref{fig:force_comm}  and the 
expected trend is confirmed: The more discontinuous the adiabatic solution, the 
larger the kinetic friction. It is important to emphasize that the exponent 
$\beta$ (as determined for sufficiently small $v_0$) only depends on the 
sign of $V_1$ but not on its precise value as long as $\mid V_1 \mid$ is 
not too large.

For 1-dimensional, inc. surfaces, the basic picture is similar. If 
$V_1$ is larger than a (positive) critical value $V_1^*$, whose precise value 
depends on the lattice mismatch, then pops between inequivalent positions are 
present and $F_k$ remains finite as $v_0$ tends to zero. For $V_1 < V_1^*$, 
however, the time derivative of the (continuous) adiabatic solution remains 
finite at all times. Hence the microscopic (peak) velocities $v_p$ scale 
linearly with $v_0$, which implies Stokes-type friction in that regime 
(1-d, inc.). Again, the exponent $\beta$ only depends on the sign of 
$V_1-V_1^*$. More detailed results including an analysis of significantly
more comples systems will be presented in a separate 
paper~\cite{aichele02}.

There is however an important difference between com. and inc. systems, when 
the analysis is extended to two-dimensional interfaces: In 2-d inc. systems, 
first-order instabilities can now occur without higher harmonics. The reason is 
that in 2-d, atoms can circumnavigate the points of maximum longitudinal force.
In 2-d com. surfaces, the behavior remains qualitatively similar as in 1-d, 
because the interference of $V_t$ and $V_b$ remains similar. Large-scale 
molecular dynamics simulations by He and Robbins support this argument even 
if the boundary lubricant is not any longer in the impurity regime. They find
small $F_k$ and large $F_s$ between com. surfaces, while no such gap is
observed for inc. systems~\cite{he01}. 
Also experimentally, signs for the effects of
increased static friction of com. surfaces in the presence of
a lubricant were reported~\cite{ruths00}.

\begin{figure}[hbtp]
\begin{center} \leavevmode
 \hbox{ \epsfxsize=90mm
 \epsfbox{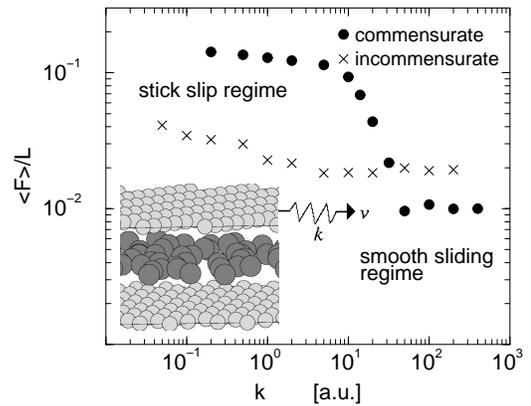} } \vspace*{-5.3cm}

 \begin{minipage}{8.5cm}
 \caption{Kinetic friction force $F_k$ devided by load $L$
 as a function  of spring constant $k$ for com. and inc. walls
 lubricated by a quarter layer. 
 A schematic of the simulation is shown as well.
 }
 \label{fig:f_aver}
 \end{minipage}
\end{center}
\end{figure}

This has potentially measurable implications for the transition from stick-slip
motion to smooth sliding. In the stick-slip regime, friction is dominated by
$F_s$, while in the smooth sliding regime, only $F_k$ is relevant. We have
extended previous simulations~\cite{muser01}
to support this point further.
For this purpose 
the same model has been employed as that in Ref.~\cite{he99}. 
A schematic of the simulation is shown in Fig.~\ref{fig:f_aver} 
together with the average friction, as defined by 
the energy dissipated per slid distance. The verification of this prediction
requires smooth surfaces, because rough surfaces automatically
lead to inhomogeneous energy landscapes~\cite{gao00}. The data shown in 
Fig.~\ref{fig:f_aver} was produced at a normal pressure of
0.4~GPa and a velocity of about 1~m/s using
the same conversion of units as in Ref.~\cite{he99}. 
Note that the exceedingly small inertia of the slider as compared
to experiment moves the transition from stick-slip to smooth sliding 
to large velocities.

We will now turn to the discussion of the effects of thermal fluctuations. He 
and Robbins found that velocity 
dependent corrections in a 2-dimensional, lubricated, inc. interface 
satisfy~\cite{he01}:
\begin{equation}
F_k(v_0) = F_k(v_{\rm ref}) + {\cal O}(\ln( v_0/v_{\rm ref}))
\label{eq:fk_coulomb}
\end{equation}
over several orders of magnitude in $v_0$. This is different from the behavior 
found in the Prandtl Tomlinson model. Although similar temperature corrections 
have been suggested, for instance by Prandtl himself, more rigorous treatments 
yield  corrections of order $(T \ln v)^{2/3}$~\cite{sang01}. 
While these corrections describe 
atomic force microscope experiments of nanoscale single-asperity contacts 
fairly accurately, it seems that they cannot provide an explanation of 
the usually observed $\ln v_0$ corrections in a straightforward manner. 
\vspace*{-2cm}

Here it will be shown that simple logarithmic corrections are obeyed even in 
the impurity limit. Moreover, the crossover to linear response at extremely 
small sliding velocities will be included in the discussion. Fig. 3 shows the 
normalized friction force obtained at thermal energy $T = 0.07$ for the 
com. $V_1 > 0$ model. Three regimes can be identified. At very 
small velocities, friction is linear in $v_0$ and one may associate this 
regime with the creep regime. At intermediate $v_0$, 
Eq.~(\ref{eq:fk_coulomb})  is rather well 
satisfied. At "large" velocities, thermal fluctuations become less relevant 
and the motion is close to that of the athermal system. The data obtained at 
different temperatures can be collapsed on a master curve. The collapse 
requires two dimensionless scaling functions $r(T)$ and $s(T)$ that both 
depend on temperature $T$ only. The collapse is done via 
$ F_k(v,T)  = r(T^*) F_k(v^*,T^*) / r(T$ with
${s(T)} \ln (v^*(T^*)/v_1) = {s(T^*)} \ln (v(T)/v_1)$
where $v_1$ is a constant, $r(T)$ is almost constant [$r(0.02) = 0.98, 
r(0.2) = 1.2$] and $s(T) \approx k_b T / \Delta E$ where $\Delta E$ can be 
interpreted as an effective (free) energy barrier.
Qualitatively similar crossover from the linear response regime
to the activated regime are observed in many other systems such
as single particles in a static periodic potential~\cite{risken79}, 
driven thermal elastic manifolds~\cite{chauve98},
and shear-thinning  fluids~\cite{granick}.

\begin{figure}[hbtp]
\begin{center} \leavevmode
 \hbox{ \epsfxsize=70mm
 \epsfbox{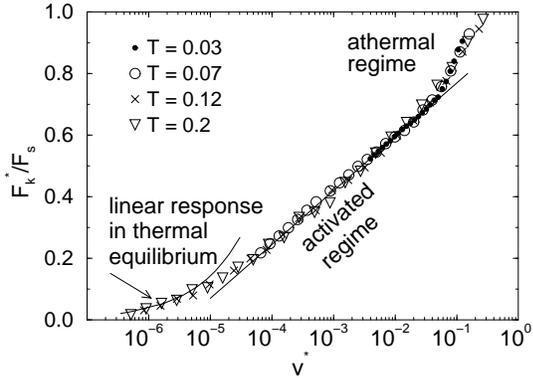} }
 \begin{minipage}{8.5cm}
 \caption{Scaling plot of kinetic friction $F_k^*(v)$ normalized by
 athermal zero-velocity  limit $F_s = F_k(v=0,T=0)$. The reference
 temperature in this plot is $T = 0.07$. In the regime of thermal
 equilibrium a linear law $F_k(v) \propto v$ is drawn to guide the eye.
 The critical $v^\beta$ contribution  is subtracted from all data.
 }
 \label{fig:scaling}
 \end{minipage}
\end{center}
\end{figure}

In conclusion, this Letter provides a classification scheme for 
instabilities that can occur when two solids are in relative sliding motion. 
First-order
instabilities are defined as pops of atoms (or other degrees of freedom)
between inequivalent positions. They lead to kinetic friction that
remains finite when the sliding velocity $v_0$ goes to zero provided
the system is athermal. The exponent $\beta$ of the velocity corrections
$v^\beta$ depends on the details of the model, however, $\beta$ only changes
its value at certain critical points in the parameter space defining
interactions and geometry. At finite temperature, corrections in the
order of $\ln v_0$ apply.
Second-order 
instabilities are defined as pops between equivalent positions. They lead 
to a sub-linear power law $F_k \propto v^\beta$. If, however, the adiabatic 
solution $x_{\rm ad}$
of the boundary lubricant moves with finite velocity at all times,
then simple Stokes friction follows. Thus an important result of this
analysis is the identification of critical points where the kinetic
friction law changes qualitatively as a parameter describing the
interactions or the geometry is varied. 

While the present study is primarily concerned with dilute boundary 
lubricants, the concept itself seems to be more general. For instance
$x_{\rm ad}$ can be a collective order parameter that fluctuates back
and forth between two values. Such quasi-periodic phase transitions 
have been observed in computer simulations of Ni asperities moving
over a Cu substrate~\cite{buldum97}. 
Of course, the situation in those simulations
was more complex, because wear occurred as a side effect of the motion.
In the other extreme, $x_{\rm ad}$ may merely denote the position of
an electronic orbital. 

I thank K. Binder for useful discussions.
Support from the BMBF through Grant 03N6015 and
from the Materialwissenschaftliche Forschungszentrum  
Rheinland-Pfalz is acknowledged.

\end{multicols}
\end{document}